**Bimodal Bias against Chinese Scientists in the American Academy: Penalties for Men, Bonuses for Women**

Gavin G. Cook

*University of Macau and Harvard University*

**ABSTRACT**

Given the recent targeting of Chinese scientists by the Department of Justice and sizable contributions of Chinese scientists to American science, it is urgent to investigate the presence and the particulars of anti-Chinese discrimination in the American academy. Across a sample of all faculty in the top 100 departments of sociology, economics, chemistry, and physics in the United States, we show that female Chinese scientists comprise a much higher percentage of the female professoriate than male Chinese scientists in the male professoriate. Using an exact matching approach, we then find that male Chinese scientists suffer from a dramatic citation penalty but that female Chinese scientists enjoy a persistent citation bonus. On average, female Chinese scientists require fewer citations on average than non-Chinese women where male Chinese scientists require more citations than their non-Chinese counterparts to attain a tenure-track professorial job of a given prestige rating.

[145 words]

# INTRODUCTION

The first incarnation of what would one day evolve into modern science emerged in Western Europe following The Enlightenment. Many have argued that science was one of the ultimate products of a 1000-year process that began with the Catholic church banning consanguineous marriage, led to the weakening of clan-based institutions, and ultimately resulted in the psychological transformation of Europe into a continent of individuals unmoored from extended families (Merton 1938; Henrich 2021). While the particular lineage of the science that is practiced around the world is not generally understood as originating in East Asia, East Asian cultures have long nurtured traditions of detailed instrumentation and empirical observation of the natural world that many have referred to as either protoscience or simply science (Needham and Wang 1988). The strong presence of an autochthonous variant of scientific or prescientific reasoning made the cultures of East Asia fertile soil for the transplantation of Western science. As a single glance at the economies of technologically advanced economics of East Asia shows, European science was indeed successfully transplanted to East Asia, and it has obviously flourished in the region, though neither science's transplantation nor eventual flourishing happened immediately. Both the fruits and techniques of science were initially shunned by elites in imperial China and imperial Japan, but they were then enthusiastically adopted hundreds of years later by new generations of the elite castes of the fledgling nation-states of both cultures beginning with post-imperial Japan and followed soon by nationalist China (Elman 2005). The nations of East Asia are now global leaders in industrial and academic science. While Chinese scientists participate very fruitfully in global science, this is not to say that Western science was transplanted in whole to China or that science developed in lockstep both in China and the lands of its origin. The path-dependent histories of empirical inquiry and institution-building in China and the foundational parameters of the cultures of China ensured that the discrete package of practices and beliefs that form the core of science was stripped of the broader cultural context in which it first developed in Europe (Elman 2005). Science took on new shapes and forms as it took root and grew in East Asia; to name one example of this, some have argued that the core institutions of Chinese science institutions, many of which have more rigid and top-down cultures than those of the Western world, have pushed Chinese science to focus more on the capital- and labor-intensive projects typical of 'big science' (de Solla Price 1963; Xie, Zhang, and Lai 2014).

While the fruits of science are often attributed to the nations to which scientists trace their descent and can confer great prestige on a nation, as was the case in the 'space race' of the 1960s between the United States and the Soviet Union (Hardesty, Eisman, and Khrushchev 2007), nation states and universities are institutions, and institutions are abstract collectives that cannot physically perform the daily embodied and often grueling labor that science demands. The frontlines of science are manned by scientists, who are necessarily human, and scientists of East Asian descent contribute greatly to knowledge production both in East Asia and abroad (Mary Paul 2022). The influence of East Asian scientific migrants is particularly strong felt in the United States. A brief glance at American luminaries in the natural sciences shows this to be the case: Fei-Fei Li, a heavyweight in artificial intelligence whose work has garnered more than 250,000 citations; MIT's Feng Zhang, one of the pioneers of CRISPR gene editing technology (Cong et al. 2013); and Michio Kaku, a quantum physicist and scientific communicator. There are, however, far fewer East Asian household names in the upper echelons of science than one would expect from the remarkable preponderance of East Asians in the lower and middle rungs of the scientific hierarchy. It is currently a fact that the conduct of science, particularly the natural sciences, in the United States is at least partially reliant on the ministrations of scientists of East Asian descent. A remarkable 88,512 Chinese students earned their PhDs in the United States from 2001 to 2020, comprising almost a quarter of all PhDs earned by international students, and many of the other top countries of origin for American PhD students are also located in East Asia, including South Korea (25,994 students) and Taiwan (12,648 students) (National Science Foundation 2022). While there is no ready-made list of the scientists that form the commanding heights of science, the lack of East Asian household names in science mirrors the under-representations of East Asians in leadership positions in corporate America (Lu, Nisbett, and Morris 2020).

While the situation of East Asian scientists in the United States is not sanguine, that of Chinese scientists is particularly dire. This is partially because Chinese Americans are tied with Japanese Americans for the longest history of continuous residence in the United States of any Asian American group, and there has been ample time for painful stereotypes about the Chinese to emerge in America and for violence to be done to Chinese America (Lew-Williams 2018). Chinese scientists are surely impacted by negative stereotypes of the Chinese as, for example, dishonest and robotic (Bui 2022). More recently, however, geopolitical tensions between the United States and China have unfortunately hampered the careers of both Chinese and Chinese American scientists in the United States. The Trump Administration's China Project has had terrible consequences for Chinese scientists in the US. The China Project and other initiatives have uncovered many cases of proven subterfuge, including but not limited to the non-Chinese Charles Lieber, who was convicted for violating the terms of NIH and DOD contracts by secretly accepting funds from the Chinese Thousand Talents Program (Office of Public Affairs 2020); Zheng Zaosong, who was arrested at Boston's Logan airport with 21 vials of biological material en route to China (Office of Public Affairs 2020); and Linwei Leon Ding, who worked for Google and tried to illegally transfer code on artificial intelligence from Google to companies in China (Office of Public Affairs 2024). Many Chinese scientists, both Chinese and non-Chinese, allege that the boundaries of what constitutes illegitimate versus legitimate exchange with China are overly capacious and that the China Project has created an atmosphere of fear (Mervis 2022; Lee and Li 2023; Xie et al. 2023). Both Chinese and non-Chinese scientists mourn the painful consequences for Chinese scientists who were falsely accused of wrongdoing and forced to spend years clearing their names, such as Feng Franklin Tao and Gang Chen, and also for the Chinese scientists who may have done no wrong but live under the threat of investigation and the nebulous specter of anti-Chinese sentiment (Mervis 2022; Mervis 2023; Mervis 2024). The FBI and DOJ under the second Trump administration have zealously sought and persecuted Chinese scientists (U.S. Attorney's Office, Eastern District of Michigan 2025a; U.S. Attorney's Office, Eastern District of Michigan 2025b), and they have also communicated these cases and their outcomes more broadly than their predecessors in previous administrations.

While it would be convenient to pin the blame for the anti-Chinese sentiment in American science on the first and second Trump administrations and right-leaning Americans, there are many other initiatives beyond the China Project that similarly create barriers, psychic or bureaucratic or otherwise, to American professors working with Chinese nationals either within or beyond the borders of the United States. MIT in particular has adopted an official stance of wariness toward collaboration with certain institutions in China. MIT's China Strategy Group[1] has recommended that the university not accept visitors or postdoctoral fellows who are affiliated with the seven military-adjacent universities that are colloquially the 'Seven Sons of National Defense' (国防七子, *guo fang qi zi*) and has cautioned faculty against participating "in 'talent recruitment' [sic] programs that are designed to transfer technology to China," such as the "thousand talents" program (MIT China Strategy Group 2022). While MIT places additional scrutiny on any collaborations with China, Saudia Arabia, and Russia (MIT Global Support Resources, n.d.), only China has sparked the convening of a country-specific strategy group. The predictable result of these policies is that research exchange between the US and China has slowed since 2019, and many Chinese scientists have even left the United States (Zhu et al. 2021; Quinn 2023). Regardless of whether one thinks the increased scrutiny on Chinese scientists is justified or not, that this scrutiny exists at all creates an urgent need for foundational research on the plight of Chinese scientists in the United States.

---

[1] The group was chaired by a high-ranking provost of international affairs and likely has considerably influence. Its affiliation with the MIT name and official MIT insignia alone conveys significant visibility and prestige.

As Chinese scientists in general suffer under anti-Chinese discrimination, so do female scientists suffer under various forms of gendered discrimination. Many researchers in race and gender have adopted an 'intersectional' framework that posits that race and gender are independent vectors of oppression that either add or multiply in the 'multiply-burdened' to produce far more dire outcomes than either vector in isolation (Crenshaw 1989). Chinese women are formally considered 'women of color' by many, and this framework would lead us to assume that Chinese women suffer from more discrimination than do Chinese men. Because the predictions from intersectionality and similar theories of discrimination have not been tested with regard to the reception of Chinese Americans in science, we will additionally investigate how anti-Chinese bias varies with gender.

To understand the presence and potential extent of bias against male and female Chinese scientists in the United States, we rely upon a unique dataset composed of bespoke and found data. We first collated a proprietary database of the CVs of all professors in the top 100 departments in the United States in four fields: sociology, economics, physics, and chemistry. This database includes information about the Chineseness and gender of every professor it covers. We then link this database to Microsoft Academic Graph, a large-scale bibliometric database with extensive coverage of both authors and publications (Sinha et al. 2015). We begin by providing summary statistics of Chinese scientists in our four fields of interest across three tiers of institutional prestige as defined by positioning on the US News and World Report (hereafter USNWR) rankings: High (top 1-15), Medium (top 15 - 50), and Low (top 50 - 100). Following many decades of sociological inquiry on inequality and stratification, we aim to detect bias by investigating how the relationship between performance and compensation varies across race in the American academy. As have many other scholars since the founding of scientometrics, which is now called 'the science of science', as a distinct field (Fortunato et al. 2018), we use citation counts as a flawed but generally reliable proxy for article quality. Overall, we find that there is a Chinese penalty for Chinese men but a Chinese bonus for Chinese women. This bimodal penalty and bonus emerges both in the distribution of placements and in the requirements necessary to attain a placement. If we quantify the requirements necessary to attain a given academic placement as the gender-, cohort-specific, and university ranking-specific citation count necessary to attain a job, we find female Chinese scholars enjoy a citation advantage relative to non-Chinese women where Chinese men suffer from a citation disadvantage relative to non-Chinese men. Our findings are at odds with many theories of how identity categories overlap to produce stratification and cumulative disadvantage, including intersectionality, and instead point to the potential benefits of interaction-based, bottom-up models for evaluating patterns of difference, division, and stratification.

**LITERATURE REVIEW**

**ASIAN AMERICANS AND RACE IN AMERICA**

The unique racial context of the United States provides the context for our investigations into bias against Chinese scientists. There are 5 official racial categories on the United States Census: White, Black, Native American/Alaska Native, Asian, and Native American/Other Pacific Islander. There is also the non-racial ethnic category of Hispanic. The interracial relations of most groups in this hierarchy are summarized clearly by theories of racial inequality in the United States, which argue that America is marred by an ugly history of 'systemic racism' and that there is a putative racial hierarchy in America with Whites at the top and African Americans at the bottom. Many volumes have been authored on this topic (Kluegel 1990; Paige and Witty 2010; Rowley and Wright 2011; Gans 2012; Misir 2022; Banaji, Fiske, and Massey 2022). Though their plights are discussed less often, the pre-Columbian denizens of North America are assumed to join African Americans at the bottoms of this hierarchy, and the relatively admixed progeny of Iberian, African, and pre-Columbian Central and South Americans who are collectively called Hispanic Americans occupy a position close to the bottom (Massey 2014; Michener and Brower 2020).

This is not, however, an open and shut case. These categories are far from perfect, and the Hispanic category is an excellent case of why this is so. Hispanic America cannot be easily slotted into the America's presumed White-Black racial system because 'Hispanics' are an admixed population with a broad range of skin tones (Perreira and Telles 2014), and even if we condition on the Hispanic groups that are sometimes considered White-adjacent, notably Cuban-

Americans, we find that there is extraordinary diversity of skin color even within nationally-bounded Latin American subgroups (Portes 1984). Recent work has shown that the advent of Latin Americans has perturbed long-standing dynamics between White and Black America (Abascal 2015, 2020). Scholars of ethnicity contend that the 'Hispanic' label is less a reflection of demographic reality than a resultant outcome of the political jousting between activists, businessmen, and unelected administrators (Mora 2014).

If Hispanic America's position in the American racial hierarchy is complicated, that of Asian Americans is more complex still and additionally very precarious. Their position is complex is because, for one, the category of 'Asian American' is remarkably capacious. While it is perhaps not realistic to expect or demand that census categories map to the complicated categories of ethnicity with any measure of fidelity, the census category of AAPI (Asian American Pacific Islander) is vague enough to be in a category of its own even among its aleady vague neighbors, the categories of White and Black. 'AAPI' subsumes almost every human east of the medieval boundaries of Europe as demarcated by Bosphorus strait and the Ural mountains and then some. This mega-region ranges from Turkey to Tuvalu and beyond to the Hawaiian archipelago, and it encompasses roughly 60 percent of the world's population. As one might expect, the groups that form the AAPI supergroup of almost 5 billion people have experienced diverging destinies in the United States. The big tent of the AAPI census category includes Chinese and Indian America, who sit at the top of America's income distribution and are far wealthier on average than Whites, and the impoverished Cambodian and Hmong communities, who cluster at the bottom end of the income distribution (Sakamoto, Goyette, and Kim 2009). The position of Asian Americans in America's racial hierarchy is precarious, then, because many Asian Americans are successful. Asian Americans are often called a 'model minority,' and scholars of Asian America have long discussed the impact and import of what they have termed the 'model minority myth' (Sakamoto, Goyette, and Kim 2009; Sakamoto, Takei, and Woo 2012; Lee and Zhou 2015).

There is a third and far subtler reason that the census category of 'Asian' is so complicated: many Americans assume that 'Asian' is shorthand for 'East Asian' and, more specifically, 'Chinese' (Lee and Ramakrishnan 2020; Goh and McCue 2021). While the remarkable heterogeneity of Asian America means that the experiences of no single Asian American group can be seen as representative of other Asian American groups, Chinese Americans have a very long history in America, and the uniqueness of the Chinese American experience is at odds with their automatic elision as prototypically 'Asian' by non-Asian Americans. The Chinese Americans are the largest constituent /group/ of Asian Americans in the United States, and they are tied with Japanese Americans for duration of residence in the United States (Lew-Williams 2018). The combination of demographic weight and historical time has created many opportunities for America to be unkind to the Chinese American community. The Chinese Exclusion Act is the most dramatic example of discrimination against the Chinese people, but it is far from the only one. The first Chinese immigrant to the United States arrived in California in the 1840s, and many other Chinese migrants followed suit. The first Chinese Americans worked in gold mines and then on railroads in back-breaking conditions (Lew-Williams 2018, Chang 2019). Most Chinese migrants to California were men, and they were seen as robotic, soulless, automaton-like workers who were pushed to inhuman extremes by their bosses (Bui 2022). The Chinese in San Francisco were maligned as sources of vice, including but not limited to prostitution, gambling, and opium, and diseases ranging from leprosy to syphilis (Trauner 1978). While the association of foreigners with disease has occurred since time immemorial and may even have an evolutionary basis, the COVID-19 pandemic unfortunately showed America that the alignment of the Chinese people with disease in the anglophone world endures to the present (Cook, Huang, and Xie 2024). Many medical professionals worried that the ills of the Chinese interacted to form vicious cycles of vice; for example, it was a common fear that opium smoking would increase the sexual desires of White men and cause them to father miscegenated offspring with Chinese women (Ahmad 2000). Concerns over the influence of the Chinese on America led to the Chinese Exclusion Act of 1882. This anti-Chinese sentiment continued into the 20th century. Explicitly anti-Chinese tirades, including the infamous *Ways that are Dark* (Townsend 1933), continued to be favorably received in the United States into the 20th century.

**INTRA-ASIAN DIFFERENCES**

Whenever Chinese Americans are compared to other Asian American groups, the uniqueness of Chinese America is immediately apparent. It is easiest to begin by comparing Chinese Americans to the American-resident diaspora communities of China's regional neighbors: Korean Americans, Japanese Americans, and potentially Vietnamese Americans. For a striking example of how China is often singled out as uniquely bad among other East Asian cultures, *Ways that are Dark*, which occupies a notorious place in history as perhaps the most stridently Sinophobic book ever written in English, ferociously denigrated China and the Chinese people but praised the Japanese people (Townsend 1933). This anecdotally suggests that the anti-Chinese racism may be more intense and qualitatively different than anti-Asian racism, and many lines of empirical evidence buttress this conclusion. A 2007 survey in California of self-reported discrimination reports that a higher percentage of Chinese Americans than Japanese

Americans reported experiencing discrimination (Gee et al. 2009). During the COVID-19 pandemic, Trump infamously labeled COVID-19 the 'China Virus,' and anti-Chinese sentiment on Twitter, now simply X, spiked dramatically in the wake of COVID-19's arrival on American shores (Cook, Huang, and Xie 2024). While the COVID-19 pandemic is not useful for directly comparing anti-Chinese sentiment with discrimination directed against other East Asian Americans, surveys of Chinese Americans after COVID-19 show consistent evidence of Chinese Americans self-reporting discrimination at higher rates than other East Asian groups (Hahm et al. 2021; McGarity-Palmer et al. 2024). Most relevantly, Chinese restaurants reported a sharper decline in patronage during the COVID-19 pandemic than other comparable restaurants, and this effect was larger in communities that voted for Trump in 2020 (Huang et al. 2023).

    Wealth is another lens of comparison. We may contrast Chinese America with the only Asian American group with a higher average yearly income than Chinese Americans: Indian Americans. While, in the aggregate, Chinese Americans are wealthier on average by far than White Americans, Chinese Americans are underrepresented at the extremes of wealth, and there is a suspicious dearth of Chinese Americans in the highest rungs of American money, privilege, and power. Many call the mechanism that prevents Chinese Americans from rising to C-suite positions in corporate America the 'bamboo ceiling,' an appropriation of the 'glass ceiling' that is purported to block the ascent of women in corporate America (Lu, Nisbet, and Morris 2020). The dearth of Chinese executives becomes most evident when Chinese Americans are compared to Indian Americans, who famously dominate the C-suite leadership of America's flagship technological corporations. To name a few high-profile examples, the CEOs of Microsoft, Alphabet (the parent company of Google), Adobe, YouTube, and IBM are all ethnically Indian.

There is a growing body of work that compares East Asians in America to South Asians in America. Some explanations for the striking differences in leadership representation between East and South Asians in America point to the relative ability of South Asians to form connections with non-coethnic alters and the relative inability of East Asians to do the same, which is to argue that South Asians are better than East Asians at networking with outgroup members (Lu 2022). This is puzzling, however, given that South Asians are rated as more ethnocentric than East Asians in some contexts (Yousaf et al. 2022). Both Lu (2022) and Yousaf et al. (2022) use an instrument called the Intercultural Willingness to Communicate (IWTC) to measure ethnocentrism, but where Lu found that East Asian students were less willing to communicate with non-East Asian students at elite American business and law schools, Yousaf et al. found that Pakistani college students in Pakistan were less willing to communicate interculturally than Chinese college students in China. Regardless of ethnocentricity, South Asians are consistently evaluated as more assertive than East Asians, which might explain their relative success over East Asians (Lu, Nisbet, and Morris 2020; Lu, Nisbet, and Morris 2022). There is a stark difference between the intensity of in-group cohesion and propensity to out-group networking, and the paucity of assertiveness among East Asians could partially explain why East Asians are not as skilled at networking with outgroup members as other minority groups. The placements of both groups on the Stereotype Content Model (SCM) tentatively supports this hypothesis: while Americans sort Indian Americans into the 'Ingroup/Allies cluster,' defined by high competence and high warmth, American slot East Asians into the stereotypically high competence but low warmth 'Competent but not Nice cluster' (Lee and Fiske 2006). We may bundle the ethnocentricity and assertiveness questions into one hypothesis, which we may call H1: the high ethnocentricity + low assertiveness hypothesis for bias against

East Asians, and this hypothesis specifically predicts that Chinese Americans may perform less well in fields with lower percentages of coethnic alters.

Despite its widespread usage, The Stereotype Content Model only captures 2 dimensions of stereotype construction, and it necessarily fails to capture other dimensions of intergroup conflict that may also be relevant for evaluating the trajectories of Asian Americans in the sciences. For example, by many measures that are not captured by their high rankings of warmth and competence in the SCM, South Asians suffer from more far discrimination than East Asians on vectors of race and religion. Loosely paralleling a growing body of work on colorism in the African American community (Monk 2015), some extant evidence suggests that darker-skinned Asian Americans suffer from a proportionally higher discriminatory burden than lighter-skinned Asian Americans (Ryabov 2016a; Ryabov 2016b). As many non-Chinese Asian Americans were misidentified as Chinese Americans and then physically attacked for the association of Chinese America and COVID-19 during the height of the COVID-19 pandemic (Grover et al. 2021), many South Asian Americans were maligned as brown and Muslim and stereotyped as terrorists in the wake of the September 11$^{th}$ attacks (Cainkar 2018). Americans additionally tend to prefer Chinese respondents over Indian respondents in canonical tests of outgroup bias. Americans would generally prefer to live next to a Chinese family over an Indian family and would rather their children marry people of Chinese descent than Indian descent (Lu, Nisbet, and Norris 2020). Despite the combined racial and religious animus they face, Indian Americans have attained a level of success that is paradoxical when seen through the lens of 'colorism' and the American racial hierarchy.

Taken together, this evidence suggests that the drivers of bias formation regarding Americans of Chinese and Indian descent are very different. The complexity inherent in entangling these drivers leads us to address Indian America in future work. This evidence also implies that naked racial hatred on the basis of skin tone or racialized appearance is not fully responsible and likely not primarily responsible for the putative bias against Chinese Americans. We must instead look to other areas for the font and the bulk of the prejudice against Chinese America.

**THE SUCCESS OF CHINESE AMERICA: CULTURE OR SELECTIVITY?**

The origins of the remarkable success of Chinese America are a matter of long-standing debate in the social sciences, and the sociological literatures on Asian America have converged on two mutually contradictory explanations for it. The first is the 'hyper-selectivity' hypothesis (Lee and Zhou 2015), which argues that Asian American migrants are often from high-status backgrounds and that this selection effect is the primary cause of the successes of Asian America. The evidence for this hypothesis, however, is mixed. The Chinese migrants to California in the 1800s were generally from the lower classes, and, despite their meager backgrounds, Chinese and Japanese Americans had equaled or exceeded the educational attainment of White Americans by 1930[2] (Hirschman and Wong 1986; Tian 2023). Asian Americans additionally reached economic equality with White Americans by the 1970s (Hirschman and Wong 1984). The second explanation may be termed the cultural hypothesis or the overachievement hypothesis. It posits that Asian American success in education is due to non-cognitive factors, particularly a culture-wide emphasis on academic success (Hsin and Xie 2014). Many Asian American cultures also link effort and achievement very explicitly (Hsin and Xie 2014). Other cultures in the United States view the link function between the two differently, and some scholars have suggested that the many cultures of African-America view racism as blocking the returns from academic effort and instead attribute racial achievement gaps to structural factors (Bañales et al. 2020). It is one thing if a child's parents reward the child in the family for academic achievement and another if the child's success redounds to his or her community, who then reward the child in turn. In the case of Chinese Americans, both are true (Fuligni 2001; Lee and Zhou 2015). Present in both frameworks is the idea that Asian Americans may feel a need, whether rationally or irrationally, to out-compete their non-Asian alters in the job market due to perceived discrimination. As Asian American families spend more on education than White parents (Tian 2023), so may Asian Americans feel a need to acquire advanced educational credentials to compete in a job market in which they are both "presumed competent" and the targets of discrimination (Lee et al. 2024).

---

[2] It merits mentioning that Hirschman and Wong, who were staunch opponents of the cultural hypothesis, found that Chinese Americans were doing better than most White Americans on some measures of academic success in the 1930s, a full thirty years before the beginning of mass migration and fifty years after the Chinese Exclusion Act.

**RACE, SEX, AND INTERSECTIONALITY**

Gender and racial stereotypes do not exist in isolation, and many racial stereotypes are implicitly gendered. The prototypical example of this association is that people of African ancestry are seen as more modally masculine than Whites where those of East Asian ancestry are seen as more feminine on average than Whites (Galinksy, Hall, and Cuddy 2013). The most famous framework for exploring the intersection of race and gender is fittingly termed intersectionality. First outlined by Kimberle Crenshaw, the framework argues that non-White women are "multiply-burdened" and urges scholars to "contrast the multidimenstionality of Black women's experience with the single-axis analysis" of White women and Black men (Crenshaw 1989, 139). While Crenshaw does not mention the plight of Asian or Hispanic

Americans of any stripe, others have applied Crenshaw's work to these groups (Ghavami and Peplau 2013; Rosette et al. 2016). The verbiage Crenshaw employs brings to mind linear algebraic framings of matrices and vectors and implies that racism and sexism are independent vectors of oppression. Crenshaw adds that "the intersectional experience is greater than the sum of racism and sexism," and while this means that racism and sexism cannot be summed directly as 'racism + sexism', it implies that their relationship is fundamentally additive (Crenshaw 1989, 140). One way of interpreting this is that overall oppression burden of racism and sexism is the sum of racism and sexism plus an additional factor or the product of race multiplied by sexism. The full explication of Crenshaw's framing of is: "To bring this back to a non-metaphorical level, I am suggesting that Black women can experience discrimination in ways that are both similar to and different from those experienced by white women and Black men. Black women sometimes experience discrimination in ways similar to white women's experiences; sometimes they share very similar experiences with Black men. Yet often they experience double-discrimination - the combined effects of practices which discriminate on the basis of race, and on the basis of sex. And sometimes, they experience discrimination as Black women - not the sum of race and sex discrimination, but as Black women" (Crenshaw 1989, 149). Whether the oppression that Black women is experience is an additive blend of their experiences as women and non-White or a wholly unique third option, the end result is that Black women are ultimately more oppressed than either White women or Black men. Others have operationalized this approach along these lines and have argued either for the uniqueness of the Black or non-White female experience (Ghavami and Peplau 2013; Billups et al. 2022) or that racism and sexism can be summed (Pogrebna et al. 2024). Some explicitly argue for an "additive model" (Liu and Wong 2018) of discrimination or a "double jeopardy" or "double disadvantage" hypothesis (Beal 2008; Denise

2012; Denise 2014). Regardless of how the vectors of oppression corresponding to race or gender are taken to interact or perhaps not interact with one another, all of the above theories contend that Black women experience more oppression than White women or Black men individually. Recent guides on intersectional science and empirical work on how the public perceives intersectionality in science support the idea that, however one adds these variables, they always end up adding to the detriment of the multiply-burdened, arguing that Black women experience worse outcomes than White women or Black men (Eom et al. 2025; Nielsen et al. 2025). We may refer to this as H2: the intersectional hypothesis.

A weakness of the literature on the diverging destinies of Asian American groups is that it sometimes does not differentiate between the trajectories of Asian men and Asian women in America. Given the wide variance in stereotypes attributed to Asian American men and Asian American women, it would stand to reason that American culture would receive and perceive Asian Americans of both sexes differently. Intersectionality would predict that the bamboo ceiling and glass ceiling interact in some way, which would lead to Asian women suffering from harsher professional biases than Chinese men. Extant research in the intersectionality framework on the experiences of Asian women in the US and the UK argues that the two ceilings do, in fact, interact and that Asian women are burdened by both racism and sexism (Lane, Tribe, and Hui 2011; Ching et al. 2018; Forbes, Yang, and Lim 2023). A recent review by Lee, Goyette, Song, and Xie suggests that there a 'double disadvantage' for Asian American women in advancing to leadership roles (Lee et al. 2024). Some evidence suggests that the increasing femininity attributed to Asians in general stacks additively with the femininity that Asian women possess by virtue of their gender, and this doubled femininity renders them less suitable than non-Chinese women in male-typical roles in the eyes of other Americans (Alt et al. 2024). This mirrors some parts of the 'double disadvantage' hypothesis. We might call this the 'doubled femininity' hypothesis, and it would lead us to predict that Asian women suffer more than non-Asian women in fields that are seen as prototypically masculine, such as physics (Gonsalves, Danielsson, and Pettersson 2016). We label this H2a: the doubled femininity hypothesis as an addendum to H2: the intersectional hypothesis.

There is a large body of work about the plight of female scientists (Xie and Shauman 2009). It suffices to note for our purposes that many have argued that women in science face barriers that their male counterparts do not and that these barriers create intense stratification. Female scientists may publish similar amounts of papers and are professionally rewarded at similar rates for these publications, but they drop out of academia far more often than do men and so publish fewer total papers over careers than do men (Huang et al. 2020). This higher drop-out rate is understandable, for science can be hostile to women. Some teams have found that women in science are less likely to be rewarded with authorship because their work is not recognized (Ross et al. 2022). In some universities, it is hypothesized that professorial recruitment procedures may even prevent women from finding success (Nielsen 2016). A recent causal design suggests that female physicists receive a greater bump to visibility than do male physicists upon election to the National Academy of Sciences, implying that the work of female physicists is consistently undervalued before election to the NAS (Li, Zheng, and Clauset 2025).

**STEREOTYPES AS A POSSIBLE MECHANISM FOR ANTI-CHINESE BIAS**

We have established that simple racial animus is not entirely to blame for any bias against Chinese scholars, for non-Chinese Americans view Chinese America relatively favorably. What, then, might drive potential bias against Chinese American scholars? One possible set of mechanisms for this bias may be found in the bevy of stereotypes that Chinese Americans face. Due to the length of contact between Chinese America and non-Chinese America, there is a relatively rich set of cultural templates in America that may be used to talk and think about Chinese Americans. In the verbiage of the Stereotype Content Model, East Asians are generally seen as high in competence but low in warmth, which generally leads to resentment and envy (Lin et al. 2005). This is a serviceable heuristic, but science is a complex pursuit, and the two-dimensional combination of warmth and competence may not be adequate to describe how America evaluates the participation of Chinese America in science. Though empirical evidence on many of the more common but granular stereotypes of Chinese Americans is lacking, they merit explanation because the bulk of these stereotypes are immediately relevant to professional advancement in the sciences.

**CHINESE AS ROBOTS AND CHEATS**

Competition, particularly scholarly competition, is another major arena of stereotypes of the Chines people. Are Chinese Americans ruthless businessmen whose children 'ruin' school districts (Warikoo 2022), as elite Whites in of the suburbs San Francisco, who pull their children out of public schools when Asian Americans enroll in large numbers, seem to assume today (Boustan, Cai, and Tseng 2024)? If the Chinese in the United States are seen as needlessly competitive automatons, a stereotype that has a particularly strong valence in academic settings, then they may encounter discrimination when applying to educational institutions. Whether or not this is the case and, more specifically, whether or not affirmative action has negatively impacted the fates of Chinese Americans in college admissions has been litigated extensively over the past decade and resulted in the 2023 decision banning affirmative action in the United States (Arcidiacono, Kinsler, and Ransom 2023). A series of foundational works in the sociology of science (Long 1978; Long, Allison, and McGinnis 1979) argues that one of the best predictors of professional success in the academy is admission to a prestigious PhD program. If Chinese Americans are excluded from PhD admissions, the downstream consequences that would follow from this discrimination would include a resulting underrepresentation of Chinese scholars in the professoriate. There is no evidence for the underrepresentation of Chinese PhD students in PhD programs, though there is evidence of possible overrepresentation (National Science Foundation 2022). A full exploration of this question is beyond the scope of this paper.

If, at a first pass, it seen as bad by Americans that the Chinese are competitive, stereotypes about the ways in which the Chinese compete with non-Chinese may worsen how non-Chinese Americans perceive this negatively-valenced competitiveness. A neighboring stereotype to the Chinese as competitive is that of the Chinese as cunning, cutthroat, and dishonest. This image thrived throughout the 20th century in anti-Chinese pamphlets and in popular media and was most directly embodied in Fu Manchu, a prototypical Chinese villain seen as an avatar of the Chinese threat to the American people (Frayling 2014). While there is little direct academic work on how the world perceives the purported honesty of the Chinese people, there is some work on honesty of Chinese people in China as measured by coin-flip test (Hugh-Jones 2016) or reporting lost wallets as stolen (Cohn et al. 2019; Yang et al. 2023), though the results and experimental design of the lost wallet experiments in particular have prompted fervent debate (Hung et al. 2025). In short, summaries of extant evidence have so far concluded that honesty operates differently across cultural contexts without arriving at absolute measures of honesty that are fungible across cultural boundaries (Blum 2007).

Any stereotype of Chinese scientists either in the West or in China as inveterately dishonest would likely have profound and dramatic effects on the careers of Chinese scientists. This is because science prizes itself as impartial, and the singular deadly sin of science is fraud. Merton did not list honesty as one of the four core values of science, and while this remission may be seen as an error given the steady advance of fraud in the past 50 years since his writing on these norms, the Mertonian norm of disinterestedness could be read as an imperfect analogue for honesty (Merton 1974). The main values of American scientists in the latter half of the 20$^{th}$ century in their words were "objectivity, honesty, integrity" (Hajek, Paul, and Ten Hagen 2024), and the same is broadly true of science before the 1900s (Engberts 2022) and into the present (Shapin 2008). English-language discourse on the honesty of Chinese scientists has been tainted by discriminatory overtones. As recently as 2006, a column in *Science* described China as a 'scientific Wild East' with "an unprecedented number of researchers... accused of cheating - from fudging resumes to fabricating data - to gain fame or plum positions" (Xin 2006). Whether these attitudes have changed in the intervening 20 years is an open question. Recent work has focused on the propensity of the incentive structure of Chinese science at best fails to punish fraud and at worst incentivizes it (Lei and Zhang 2018). It is difficult to predict how stereotypes of Chinese scientists in China might carry over to those about Chinese scientists in America, but the nature of discrimination and the global nature of science suggest that the operative stereotypes for one group of Chinese scientists likely apply to all Chinese scientists.

It is similarly difficult to predict how stereotypes of dishonesty both scientific and otherwise would affect the careers of Chinese American scientists. As a first pass, we might predict that the careers of Chinese scientists would suffer the most in fields with the most problems with fraud. To establish which fields are most prone to fraud, it is possible to look to replication or reproducibility rates by field, but this is not a very useful metric because a vanishingly small percentage of published papers are replication studies[3]. Additionally, replication rates do not vary much across fields (Gordon et al. 2020). The finding that only 39% of psychology studies replicated famously began the 'replication crisis' in psychology in 2015 (Open Science Collaboration 2015), but the corresponding figure for cancer biology, widely considered to be a more rigorous and 'harder' science, is the equally dismal 37% (Errington et al. 2021). Finally, a paper's lack of ability to replicate does not necessarily imply fraud. Retractions, however, are a stronger but still imperfect but signal of fraud, and we may indirectly operationalize the prevalence of fraud by comparing retraction rates across field. Data on this is unfortunately outdated but suggests that rate of retractions is 0.3% in economics, 0.1% in sociology, and as high as 0.2% in some physics subfields and 0.4% in some chemistry subfields (Grieneisen and Zhang 2012, Table S1). The higher presence of fraud as measured by retraction in the natural sciences would lead us to predict that, if stereotypes of dishonesty hampered the careers of Chinese American scholars, the influence of these stereotypes would be strongest in the fraud-prone and hence presumably fraud-wary natural sciences. We may label this hypothesis H3: the dishonesty stereotype and fraud guarding hypothesis.

**STEREOTYPES OF CREATIVITY: ASIANS AS ROBOTS**

If this stereotype of Asians as robotic is taken to the extreme, the work of Chinese scholars may be seen as derivative and unoriginal regardless of how original it might be. The ideal situation to test this hypothesis would be a more rigorous variation of the classic audit study

---

[3] 0.1% of papers in economics are replication studies (Mueller-Langer et al. 2019).

where rotating panels of professors could be asked to evaluate the work of hypothetical White, Black, and Asian scholars. The work of each scholar would also rotate, and every panel would be exposed to a different configuration of race and scholarship. Though unobservable variables would muddy any causal inferences that might be drawn from the result, we would then be able to partially isolate how race affects the perception of scholarship. A prototypical small-scale natural experiment of this sort has already occurred and was described in Lu (2024). The violinist Joyce Koh related how the classical music critic Tom Deacon described the exact same recording as "beautiful" when he thought it was performed by a White woman and "faceless... neat as a pin but utterly flaccid" when he thought it was performed by an Asian woman (Koh 2021). This might apply to subfields in the social sciences. Work on how sociologists view specific subfields in sociology is scarce, and what does exist is almost exclusively devoted to the anxiety that sociology undergraduates feel towards learning statistical methods (Williams et al. 2008; Ralston 2020). We are concerned with how professors might view the work of other academics. Given the stereotype that Asians excel at math, one could imagine that methods-related work might be seen as ground-breaking if authored by a non-Chinese scholar but derivative and rote if authored by a Chinese scholar.

  This devaluation effect might apply between as well as within disciplines. The anecdote of Deacon judging the same piece differently based on perceived race of performer is certainly damning, but it is from the intensely subjective art world. To extend Merton and Zuckerman's work on scientific consensus (Merton 1974), it is possible to think of the arts as an extremely soft science where there is no consensus and no shared understanding of essential truth. This would then mean that perception and prestige signals form the main basis for evaluating excellence in art. If we assume that the arts are soft sciences and also that the anecdote of anti-Asian racism in classical music applies universally to all arts, both of which are very grand assumptions, it would imply that perception matters less in the soft sciences than in the arts and less in the hard sciences than the soft. If we follow this operationalization of the creativity hypothesis, we would expect to see more bias against Asians in the soft sciences than in the hard sciences. We may label this H4: the creativity stereotype hypothesis.

Other scholars have provided ample evidence that Asians are stereotyped as not creative (Lu 2023). While some may argue that China's scientific productivity over the past decade puts these stereotypes to rest, many have argued that productivity and creativity are not the same thing (Simonton 1997). There is very little direct work on the question of how creative Chinese scientists are in practice. Work on creativity in Chinese cultural contexts has largely analyzed the putative creativity of Chinese people in predominantly Chinese societies (Fan 2014; Cheung and Yue 2019; Wu and Albanese 2014) and has generally avoided comparisons between greater China and the non-Chinese world with very exceptions, one of which may be found in Tang, Baer, and Kaufman (2015). Recent meta-analyses have shown that the gap in creativity between 'Western and Eastern cultures,' which currently favors the West, has narrowed in younger cohorts (Barth and Stadtmann 2024). While it could be argued that the creativity of any scientific work is only tangentially or even orthogonally related to its empirical value, creativity has long been prized across the hard and soft sciences (Rothenberg 1979; Simonton 2004). To call a fellow scientist's work 'uncreative' might be a way to cast aspersion without addressing the import or validity of the work's conclusions. It is, in other words, a serious charge as well as an insult that is acceptable in scientific circles, which means that we would expect to see stereotypes of non-creativity to carry the most weight in sciences with lower degrees of internal consensus, which is to say the softer sciences. If the creativity hypothesis is true, we would expect to see a bias against Chinese scholars in the soft sciences.

**SEX, GENDER, AND ATTRACTION**

Sex and gender matter as well, for there are a number of gender-specific Chinese stereotypes. Are Chinese women submissive and passive, as they were commonly portrayed in the late 1990s (Uchida 1998; Prasso 2006), or are they sexual predators out to ruin the lives of young White boys (Matsubara 2003)? Are Chinese men effete, 'creepy,' feckless, and robotic, or are they as serious of a threat to White women as Chinese women are to White men (Lee 2017)? Of particular interest is how stereotypes regarding race and gender interact for East Asians. East Asian faces are generally viewed as more feminine than faces of other races of the same gender (Stephen et al. 2018), and East Asian women are viewed as prototypically feminine (Alt et al. 2024). This broad attribution of femininity may have field-specific consequences for sorting Chinese scholars preferentially into more 'feminine' fields. While there is evidence on how the 'hard' sciences are viewed as masculine (Gonsalves, Danielsson, Pettersson 2016), there is none on how or if the soft sciences may be viewed as feminine.

**ATTRACTIVENESS AND GENDERED STEREOTYPES OF ASIAN AMERICA**

Even the gender-based stereotypes of the Chinese in America that are not directly relevant to science, such as that of Chinese men as unattractive, may influence professional success via the halo effect (Thorndike 1920). If Chinese women are seen as attractive and Chinese men as sexually undesirable, then it could be the case that the perceived attractiveness of

Asian women and perceived unattractiveness of Asian men influences their respective placements on the academic job market. There is a bevy of evidence that beauty is associated with better professional outcomes in general (Hellyer et al. 2023; Bortnikoa, Havranek, and Irsova 2024), though this evidence also indicates that beauty and productivity are correlated; that students rated as good-looking by teachers earn better grades (Mehic 2022); and even that the attractiveness of a PhD student leads to more prestigious first academic job (Liu, Lu, and Veenstra 2022; Hale, Reveg, and Rubinstein 2023; Alkusari, Gupta, and Etcoff 2024). If attraction is as important to professional success as some research suggests, then the stereotypes of Asian men as less desirable are not only psychologically damaging but also potentially career-stymying or even career-threatening. We will refer to this as H5: the attractiveness hypothesis.

East Asian men suffer from a particularly intense stereotype burden. They are stereotypically seen as physically unattractive, sexually inadequate, socially awkward, and feminine, and they are aware that others seem them thusly, which has profound negative consequences for the mental health of Asian American men (Iwamoto, Liao, and Liu 2010; Wong et al. 2012). Chinese men are consistently seen as less masculine and less athletic than White men, who are in turn seen as less masculine and athletic than Black men (Wong, Horn, and Chen 2013). On the surface, this contradicts the dictates of intersectionality; should not East Asian women suffer more than East Asian men by virtue of their being women? Scholars of East Asian masculinity have offered a few theoretical approaches that may this apparent tension. The Subordinate Male Target Hypothesis argues that low-status men are the target of oppression from high-status males, and, because many scholars see non-White men occupying lower status positions than White men, this model predicts that Asian men might experience more discrimination than Asian women. (Yoo, Steger, and Lee 2010; Buhl et al. 2024). If this hypothesis were true, we would predict that Asian men would suffer the most discrimination in male-dominated fields. This is H6: the subordinate target male hypothesis.

We are left with six total hypotheses. Table 1 provides a comparison of these hypotheses and their predictions.

## MEASURES AND METHODS

### DATA

We use a combination of CV data and large bibliometric databases to examine bias against Chinese American scholars using a metrics-based approach. Inspired by the methodology used to compare racial groups and college admission likelihood in Espenshade and Radford (2009) and the index-based approach to stratification in Greenman and Xie (2008), we construct an index comparing the productivity of Chinese and non-Chinese scholars. We specifically compare both citations and papers published. We find that, on average, Chinese scholars must publish more and be cited more often to attain positions similar to those of their non-Chinese peers, but this is only true on average and in the aggregate. We then stratify our analyses on gender, prestige of host institution, and field.

### DATA

We use a combination of CV data and large bibliometric databases to examine bias against Chinese American scholars using a metrics-based approach. We begin with a database of transcribed information from CVs from the professoriate of top 100 departments across four fields: sociology, economics, chemistry, and physics. This provides information on everything that one could gleamed from a *curriculum vitae*: education history, professorial rank, employment history, and more. We then link this proprietary database to the Microsoft Academic Graph (Sinha et al. 2015) database, which contains comprehensive information on published papers, including paper-level citations. To link the two datasets, we use a combination of rule-based methods to ensure, for example, that the years of an author's published papers and PhD attainment align, and we word vectorization techniques to disambiguate authors with the same or similar names that study different topics.

The end result of the data linkage process provides a comprehensive view of every professor in the top 100 departments across four fields: sociology, economics, physics, and chemistry. For every professor in our sample, we have information on Chineseness, gender, professorial status (assistant, associate, or full professor), PhD origin, and complete publishing record including all citations for all papers.

**PRESTIGE**

We segment professorial prestige into 3 bands: Top 15, Top 15-50, and Top 50+. These bands roughly mirror how academics discuss the relative prestige of different graduate programs. For an illustration of how these clusters read in real terms to sociologists, the Top 1-15 cluster includes Stanford (tied for #1 in a fivefold tie), Northwestern University (#7), and the University of Chicago (#7); the Top 15-50 cluster includes Pennsylvania State (#24), University of Massachusetts-Amherst (#24), and the University of Arizona (#26); and the Top 50+ cluster includes the University of Florida (#64), Bowling Green State University (tied for #64), and Case Western Reserve University (#91) (US News & World Report 2025). There may be a possible confounding of field-level prestige with university-level prestige. In sociology, for example, Yale University is ranked #19, but Yale University is considered one of the most prestigious in the world by dint of its association with the upper rungs of the Ivy League. To disentangle these warring prestige signals is beyond the scope of this paper, but it is worth noting.

**INDEX-BASED APPROACH**

The main independent variable of interest is citation count. We are interested in the presence of citation disparities between Chinese and non-Chinese professors of comparable status. Following the overachievement hypothesis, we ask: does a Chinese academic need to garner more citations to his or her work to attain a job that a non-Chinese academic could attain with lesser qualifications? This is, however, not a straightforward question because of the nature of citation counts. Citation counts follow a power law distribution[4] (Peterson, Pressé, and Dill 2010), and many methodological challenges arise when performing basic operations both within and between power distributions. We first assume that the citation counts of each field are independently distributed power law distributions. Following Greenman and Xie (2008), we construct a cohort- and field-specific measure of relative citation that we term the citation multiple. While our results are robust to comparing the citations of Chinese and non-Chinese professors multiplicatively and additively, ie. via the citation multiple and then without correction, the results from the citation multiple are much more straightforward to interpret, and we present these in the main text.

---
[4] Many other aspects of academic output similarly follow a power law distribution. One of the earliest findings in the sociology of science was that author-level productivity, as defined by total production of papers, may be modeled with a power law distribution (Lotka 1926).

While citation count, which is a proxy for article quality, is quantitatively different across fields, other measures that we might use to capture the quality of any given publication instead different qualitatively across fields and would render inter-field comparisons impossible. The most important dimension of publication success that we have not explored is publication prestige, or the relative prestige of a journal in which any given paper is published, but most methods of quantifying publication prestige are not portable between disciplines. Economists are very familiar with the 'Tyranny of the Top Five' (Heckman and Moktan 2020), wherein any publication in economics' top five journals carries outsize weight. Sociologists suffer under the tyranny of sociology's top 3 journals, which are the American Journal of Sociology, the American Sociological Review, and Social Forces (Abbott 1999). While sociology and economics have broadly similar publishing cultures and, more specifically, very similar incentive structures for high-impact publishing,[5] the publishing cultures of the natural sciences are very different. We use an index-based approach to measuring citations for a similar reason: the number of papers published per year in each field varies so widely between the hard and soft sciences that a direct comparison of the citations that each paper garners is useless, but this difference is purely quantitative and may be easily resolved with normalization techniques. We might also focus on productivity and paper count, but the generally held consensus in the science of science is that publication quality is more important than publication quantity for securing a prestigious job (Kaur et al. 2015; Forthman et al. 2020). Because our main outcome of interest is professorial prestige, we focus primarily on citation count, which is the most robust proxy for article quality that is also fungible between fields. While productivity and publication prestige are very also important, especially for young faculty, publication prestige is not fungible across fields.

---

[5] It is also the case in sociology and economics that publications in a top journal are more impactful for one's career if they are solo-authored instead of authored in a team. This is not the case in physics, where large teams are the norm.

We aim to advance debates about representation by comparing the relationship between professional productivity and professional success across ethnicity in the academy. In doing so, we benefit from the uniquely quantifiable nature of science. Where previous work has focused on the representation of East Asians in leadership positions in large corporations (Lu et al. 2020) and the relationship between GPA and club leadership in business schools and law schools (Lu 2022), the outcomes in these examples are not coupled as tightly to their inputs as in the case of the professoriate allotting jobs on the basis of papers. While science does not always live up to its ideals of objective meritocracy (Merton 1974), many of the inputs that are involved in calculations of job allotment and prestige are public. These inputs mostly relate to publishing: how many papers an academic has published, how prestige the venues of these papers are, and how many citations these papers have garnered are all important components of the job application process. Because the stated ideals of science mean that scientists must at least pay lip service to these standards of objectivity in hiring decisions, any deviation from these standards should invite suspicion. We directly compare Chinese women to non-Chinese women and Chinese men to non-Chinese men following other work on gendered bias in academia (Tinkler et al. 2019).

**RESULTS**

**PRESTIGE BY CHINESENESS AND GENDER**

We begin by segmenting our data by Chineseness[6] and gender and first present basic counts of professors by Chineseness and gender across the 3 prestige bands[7]. Though purely descriptive, this information may be of interest to scholars of gender and science and technology studies on its own terms. First, we have prepared tables showing the proportions of Chinese vs. non-Chinese (Table 2) and then male vs. female professors (Table 3) regardless of prestige for our 4 fields of interest. We additionally provide a table of distribution of professorial head count by prestige band in the supplementary information (Table S1).

[Table 2 here]

[Table 3 here]

We find that Chineseness and maleness scale almost linearly with the hardness of a science. Chemistry, the sole exception to this rule, is less hard than physics but has a higher proportion of Chinese professors than physics. The proportion of males by field, however, increases in perfect lockstep with scientific hardness. Some scholars of science contend that the hard sciences are seen as masculine (Gonsalves, Danielsson, Pettersson 2016). They are indeed masculine if we take the preponderance of males in chemistry and physics to indicate masculinity.

---

[6] We recognize that 'Chineseness' is an unwieldy term but stress that it is the most straightforward way of describing the state of one's being Chinese or not Chinese.

[7] Because racial prejudice does not discriminate between Chinese people raised in America and Chinese people from China, we treat all academics in our sample with Chinese ancestry as Chinese.

This descriptive finding invites us to consider why and how Chinese scholars choose their fields of study. There are exogenous and endogenous factors that presumably sort Chinese scholars into some fields over others. Following the 'doubled femininity' hypothesis (Alt et al. 2024), we might predict that Chinese women preferentially sort into more feminine-dominated fields. This is not the case. In fact, the exact opposite is true. Chinese women represent a much higher proportion of female scholars in the hard sciences than in the soft. There may also be an endogenous set of cultural forces that influences the course of Chinese Americans through academia. While there is a lesser-known Chinese cultural premium placed on earning money and enjoying a stable life of material security (Steele and Lynch 2012; Li and Hu 2022), it is very widely understood that Chinese parents have high expectations for their children's educational success regardless of familial wealth (Li and Xie 2020). Education, which is mostly studying extant knowledge, and science, which involves producing new knowledge, are very different pursuits but are conducted in the same institutions. This overlap might generally drive East Asians to preferentially become professors at higher rates than non-Chinese people. This does seem to be the case; Chinese Americans comprise less than 1% of the population of the United States but between 4% to 11% of the professoriate for our fields (Table 2). The specific contours of this cultural preference, particularly the emphasis on educational success and on high-income jobs, may separately interact to drive Chinese scholars to select careers more well-renumerated fields. This is the case for undergraduates; East Asians in general (Xie and Goyette 2003) Chinese women are more likely to select undergraduate majors that lead to highly paid professions than White women (Song and Glick 2004). This trend continues into post-graduate education, where Chinese students are far more likely to attend graduate programs in the hard sciences or in engineering that in the social sciences, and this necessarily means that fewer

Chinese and Chinese American scholars become social scientists (National Science Foundation 2022). We see that this trend extends into professorial careers upon obtaining a doctorate degree, for Chinese scholars gravitate to harder sciences over the soft.

We have seen that certain fields skew male and have a higher proportion of Chinese scholars, and we have also seen that Chineseness and maleness are very tightly correlated. We now may additionally ask if there is any relationship between the proportion of Chinese females and Chinese males in our fields of interest.

[Figure 1 here]

We find that Chinese females enjoy a large and consistent gender-specific proportional advantage in terms of head count and placement over Chinese males. While the advantage enjoyed by Chinese women varies by prestige band, and there are a handful of field-prestige band combinations wherein Chinese females are scarce relative to Chinese males, there is no field where, on average, Chinese females are less represented than Chinese males. There are, for example, proportionally almost 3 times as many Chinese females as there are Chinese males in economics. This aligns with the work of Song and Glick (2004), who found that Chinese female undergraduates preferentially select jobs that lead to highly renumerated jobs, and the study of economics at both the undergraduate and graduate level is very directly associated with jobs with high salaries. There are proportionally 1.28 times as many Chinese females as there are Chinese males in chemistry, and the corresponding figure is 1.69 for physics. These findings directly contradict the 'doubled femininity' hypothesis. In sociology, the softest and most 'feminine' field in our work, the proportional placement advantage of Chinese females over Chinese males is present but very weak, and there are proportionally the fewest Chinese females relative to non-Chinese females in sociology. For a more complete discussion of these descriptive trends, see Supplementary Figures S1 – S4.

## MATCHING RESULTS

### COVARIATES AND MATCHING

We match on professorial rank, PhD prestige, decade of PhD, and gender. We use exact matching to match on all observable covariates. Because of this, the covariates are perfectly balanced by default, though our results are robust to other matching methods (See Figure S5). Our results are additionally robust to specifying the Chinese penalty as a difference or as a ratio, and we will present both sets of results for ease of interpretation.

Figure 2 illustrates the bias against Chinese scholars in each of the fields under analysis by calculating the gap in citations needed to attain a given position between Chinese and non-Chinese scholars as a ratio, and Figure 3 calculates the same figure as a difference. Both figures show a consistent citation penalty against Chinese men in the academy and a consistent citation bonus for Chinese women in the academy. We invert the Y-axis for both figures to clearly show the direction of the bias against Chinese scholars.

[Figure 2 here]

[Figure 3 here]

A few general trends emerge. Both Chinese males and females enjoy a citation bonus relative to non-Chinese scholars in chemistry, but this is the only field for which is the case. In all other fields, penalties against Chinese men are the rule. On the whole, Chinese men are treated more equitably in the harder sciences, which also happen to be more male-dominated and more proportionally Chinese. Chinese women suffer from a slight penalty in Top 1-15 departments in both economics and sociology, but these are the only two combinations of prestige band and discipline where Chinese women suffer a penalty, and, even then, the citation penalty they do suffer is essentially negligible. The fields that are the most hostile to Chinese males, namely sociology and economics, are also the most hostile to Chinese females, but Top 51+ programs in sociology bestow a sizable advantage upon Chinese female scholars. Where male Chinese scholars in Top 51+ programs in sociology must garner almost 10 times more citations than their cohort- and gender-matched non-Chinese peers, female Chinese sociologists in the same prestige bin require only 0.02 times as many citations as their cohort- and gender-matched non-Chinese peers.

A minor but notable secondary finding is that the direction of bias across prestige bands varies by field. Chinese men suffer the most within-discipline bias in Top 51+ programs in sociology and economics relative to more highly-ranked programs in both fields but endure a harsher penalty in Top 1-15 programs in physics than in less highly-ranked physics departments.

We may consider the specifics of the differential gap between Chinese male citation penalty and the Chinese female citation bonus. Figure 4 shows the ratio of the Chinese female bonus to the Chinese male penalty, and Figure 5 presents this result as a difference. The only field in which Chinese men enjoy a bonus relative to Chinese women is in chemistry, and the magnitude of this advantage is vanishingly small. In all other fields, Chinese females enjoy a sizable and consistent advantage over Chinese males, but the magnitude of this ethnicity-specific relative advantage shrinks in tandem with scientific hardness.

[Figure 4 here]

[Figure 5 here]

**HYPOTHESES VERIFIED OR DISCARDED**

A few of the hypotheses for bias against Chinese professors emerged readily from our summary of the literature on this topic, and we will discuss how have they have fared in light of our results in turn.

We begin with the hypotheses that were discarded. Firstly, we see that H2, the intersectional hypothesis, which predicts that non-White women suffer from a greater discriminatory burden than do non-White men, fails to explain our findings. Chinese women enjoy a clear and distinct premium both in terms of overall representation in the professoriate and in qualifications needed to attain academic employment over Chinese men. This premium is dramatic and consistent. H2a, the doubled femininity hypothesis, is also discarded. Secondly, H3, the dishonesty stereotype and fraud guarding hypothesis, which predicts that Chinese scholars may be discriminated against most intensely in fields with higher proportions of retractions if Chinese scholars are seen as dishonest, is also unsupported by our findings. Fields with the highest rates of retractions feature the highest proportion of Chinese scholars and show the least bias to Chinese scholars. While it may be the case that discrimination against Chinese scholars in the hard sciences manifests in different forms that are not captured by raw placement counts or the relationship between publication outcomes and placement, we see no evidence of the bias predicted by the fraud hypothesis in any of our findings and in fact see the opposite.

Finally, H6, the subordinate male target hypothesis, which predicts that non-White men fare worse than non-White women in some contexts because they are preferentially oppressed by White men, is partially true in that Chinese men fare worse than Chinese women, but the subordinate male target hypothesis additionally requires that White men do the oppressing. There is no evidence that this is the case. The subordinate male target hypothesis predicts that non-White men would suffer unduly in male-dominated fields, and our results show precisely the opposite: Chinese men suffer the strongest devaluation in more female fields and enjoy more equitable conditions in male-dominated fields. Chinese women, on the other hand, enjoy a weak premium in male-dominated fields and a stronger premium that varies with prestige band in more female fields.

This leaves three hypotheses as potentially true. H4, the creativity hypothesis, remains viable, but we stress that we have not specifically tested its predictions. A more intensive test of the creativity hypothesis would require much more extensive approaches to measuring scientific novelty, but if we frame creativity an ancillary quality that carries undue importance in the soft sciences, then we see just what the creativity hypothesis predicts: Chinese scientists suffer the least bias in the highest-consensus - which is to say the hardest – sciences, all of which have relatively clear standards and evaluatory criteria. H5, the attractiveness hypothesis, also remains. This hypothesis would predict that Chinese women, who are seen as more attractive than Chinese men by non-Chinese alters, enjoy a professional premium relative to non-Chinese men.

The data supports this conclusion, but to conclusively prove the attractiveness hypothesis would require much finer-grained tests of how individual-level attraction mediates the link between publication record and professional success.

The final hypothesis that emerges from the literature is H1, the high ethnocentrism and low assertiveness hypothesis, which contends that East Asians are more ethnocentric than other groups and network primarily among themselves (Lu 2022). This hypothesis would predict that the ostensibly ethnocentric Chinese would thrive in more heavily Chinese fields, and it is not disproven out of hand because Chinese scholars indeed suffer less bias in fields with stronger representation of Chinese scholars. Another point supporting the ethnocentrism hypothesis is that Chinese males have been found to be more ethnocentric and generally less willing to network with non-coethnics than Chinese females in observational settings (Yousaf et al. 2022), and, as one may extrapolate from the observational evidence, Chinese males suffer less discrimination in fields with more Chinese males. As discussed above, however, this hypothesis is not immediately tenable because people of Chinese descent may appear relatively ethnocentric by most measures but are not necessarily the most ethnocentric group in the world. While we do not have the data at hand to test any of the above hypotheses, they are at least not rejected, and we leave additional inquiry regarding these hypotheses to future work.

**DISCUSSION**

After matching on a wide range of covariates, we find that Chinese women enjoy a two-pronged advantage over Chinese men in the academy. Conditional on gender, there is a higher proportion of Chinese women than Chinese men in all 4 of the fields surveyed. The magnitude of this gender gap is dramatic; there are 3 times as many Chinese women relative to non-Chinese women as there are Chinese men relative to non-Chinese men in Top 1-15 programs in chemistry. In other words, Chinese women enjoy a placement advantage relative to Chinese men.

This difference in representation is dramatic but strictly observational. We then use a battery of matching methods to control for observed covariates. Controlling for decade of PhD, professorial rank, gender, prestige of current job, and prestige of PhD origin, we compare Chinese to non-Chinese professors using a measure of relative citation that we term the citation multiple. This allows us to ask how much more cited a Chinese scholar must be relative to a non-Chinese scholar of an almost identical background to attain the same job.

We find that Chinese males, on average, suffer from a Chinese penalty relative to their non-Chinese peers. This penalty is as high as 9.42 for Top 51+ departments in sociology, which is to say that Chinese males must be cited 9.42 times as much as their non-Chinese peers in Top 51+ departments in sociology and as low as 0.1 in Top 1-15 departments in chemistry. Chemistry is the only field in which Chinese males enjoy a persistent advantage relative to non-Chinese males. The situation is the exact inverse for Chinese females; relative to non-Chinese females, Chinese female academics enjoy a citation bonus in all fields and all prestige bands except for Top 1-15 departments in sociology and economics. By directly comparing the ratios of gender-specific Chinese female advantage to gender-specific Chinese male disadvantage, we find that the only field in which Chinese females do not enjoy an overwhelming advantage relative to Chinese men is chemistry.

The hypotheses that are not supported by our results are H2, the intersectional hypothesis; H2a, the related double femininity hypothesis; H3, the dishonesty stereotype and fraud guarding hypothesis; and H6, the subordinate male target hypothesis. With the notable exception of H3, the dishonesty stereotype and fraud guarding hypothesis, the hypotheses that fail to explain our results are top-down and theory-heavy. They assume a racial order in which Whiteness and maleness always interact to create compounding oppression in all circumstances. Our results suggest that common approaches to understanding stratification in the United States along racial and gender lines are not always accurate, particularly for understudied groups outside the White-Black binary that governs much of the theory and thinking on American intergroup relations. Specifically, the priors of intersectionality, perhaps the leading theory of gender- and race-based inequality, predict that different vectors of oppression always and invariable sum and/or compound upon one another to create increased oppression for the "multiply burdened." Our work presents an additional possibility: non-White women of particular ethnic backgrounds may enjoy an advantage relative to their male co-ethnics, and sex, commonly seen as an independent source of oppression, may ameliorate the bias shown to a given ethnos and potentially even transform bias into advantage or situational privilege. While it is certainly grandiose to do so on the basis of one study, if social scientists were to construct models from the bottom-up reality of how different groups behave in isolation and in tandem as opposed to beginning from stylized presumptions of how oppression flows from top to bottom, their models would likely enjoy more predictive and explanatory power. For similar findings in this vein, we see how the income gap between Black women and White women is almost non-existent where the income gap between Black men and White men is rather large (Chetty et al. 2020).

The hypotheses that were not immediately disproven by our work, which include the H1, the ethnocentrism and assertiveness hypothesis; H4, the creativity stereotype hypothesis; and H5, the attractiveness hypothesis, are generally only cited as explanations for inequality or stratification in domain-specific specialist literature (Lu 2022; Hale, Reveg, and Rubinstein

2023; Lu 2023; Alkusari, Gupta, and Etcoff 2024). The main factor these three hypotheses have in common, however, is that they are bottom-up theories of stratification. With the possible exception of the attractiveness hypothesis, which can integrate neatly with intersectionality and the rich literatures on the study of race-based discrimination in romantic behavior, these 3 hypotheses offer testable predictions about how a group will be received based on the characteristics of the group itself, the groups of alters with which it may interact, and the extant relationships between all of the above groups. Bottom-up theories of discrimination are particularly amenable to simulation-based methods and integrate seamlessly into a wide variety of methodological approaches, particularly methodological mixtures that integrate survey data, interviews, and experimental manipulation either via surveys or lab settings. An interactional approach to discrimination may yield more actionable insights for the remediation of marginalization and oppression for racialized, non-White groups than static, assumption-heavy models of injustice. It must be stressed that this is not to diminish injustice past and present. It is instead to say that injustice may be operationalized and parameterized more effectively with a primary emphasis on the reality of how discrimination actually operates.

Our work is not without limitations. We have not, for example, examined bias against non-Chinese East Asians or South Asians, though it should be noted that the total percentage of Japanese and Korean scholars does not appear to constitute more than 10% of the total number of Chinese scholars for any of the 4 fields in our sample. In addition to the theoretical reasons that that the Chinese experience in America is distinct from that of other Asian America groups, the numerical preponderance of Chinese American scholars among East Asian scholars provides additional justification for a focus on the experience of Chinese scholars in isolation. We have additionally listed the difficulties of further testing conclusively test the attractiveness hypothesis, the creativity hypothesis, or the ethnocentricity hypothesis for bias against East Asians in the academy, and we leave a fuller exploration of these themes for future work.

A few additional conclusions specific to the study of science emerge from our work. For sociologists of science, our results suggest that the underlying parameter of scientific hardness is useful for investigating inequality in the sciences. We find that the hard sciences are more receptive to Chinese scholars than the soft sciences. These results are particularly striking because, by almost all extant measures, the hard sciences are more likely to be openly anti-Chinese than the soft sciences. Professors in the hard sciences tend to be more conservative than those in the soft sciences (Gross and Fosse 2012). Chinese professors in the natural sciences, engineering, and computer science, all of which are generally 'hard' sciences, are about 1.5 times more fearful of being investigated by the Trump DOJ and of being targeted by anti-Asian violence than Chinese academics in the social sciences (Xie et al. 2023, Tables S10 and S11). Despite these fears, our work shows that there are proportionally more Chinese scientists in the hard sciences than the soft sciences, and Chinese scientists are also treated more equitably in the hard sciences than the soft as measured by the citation multiple. Presumably, the pull of the hard sciences outweighs the push of their anti-Asian elements, but this does not explain the relative lack of bias shown to Asians in the natural sciences. What, then, is going on? We do not have the data at hand necessary to test why this might be so, but a few tentative hypotheses merit discussion. The hard sciences may be more meritocratic than the soft sciences because the increased consensus in the hard sciences over the soft renders evaluation in the hard sciences more transparent and straightforward (Merton 1974; Hargens 1988). This would presumably diminish the role of non-cognitive factors like assertiveness in success in the hard sciences because one's work can speak for itself in high consensus fields. If Lu (2022) is correct in arguing that assertiveness partially explains the advantage of South Asians over East Asians, then East Asians might preferentially filter into fields where they feel less disadvantaged.

      The other major implication of our work for scholars of science involves reversing the direction of our analysis. We have so far discussed what the reception of Chinese scientists in various sciences can tell us about Chinese scientists. While we have asked how underlying cultural parameters might guide Chinese scientists to preferentially select certain fields over others, we have devoted most of our time to analyzing how this informs us of the American treatment of Chinese Americans. We may instead ask what the reception of Chinese scientists in various sciences tells us about those sciences. That 'hardness' may be quantified as a cline with sociology at one end and physics at the other and that the percentage of Chinese scientists in a given field is very closely correlated with hardness implies that there may be a deeper link between scientific hardness and receptiveness to Chinese scholars or, more broadly, to non-American scholars. If we assume every branch of science is a functionally identical to the other branches of science, then this finding is neither explicable nor important. If we instead assume, however, that every discipline maintains its own disciplinary culture and these disciplinary cultures vary with respect to regular variables, then this finding is both readily explicable and potentially important for understanding the globalization of science. If the social sciences are more hostile to certain types of scholars or certain types of international collaboration, then social scientists may have additional work to do in dismantling some of the obstacles that create these hostilities.

# TABLES

| Hypothesis | Prediction |
|---|---|
| H1: High ethnocentricity + low assertiveness hypothesis | Predicts that Chinese Americans will have worse career outcomes in fields with fewer Chinese scholars |
| H2: Intersectional hypothesis | Predicts that Chinese American women will suffer more from bias than Chinese American men because 'Chineseness' and 'Female' are discrete vectors of bias that intensify one another. Whether they intensify as 'Chinese' + 'Female' or as 'Chinese' * 'Female' or as 'Chinese' + 'Female' + 'Chinese * Female' is not explicated by this hypothesis, but the general prediction is that Chinese women will suffer more than Chinese men in almost all circumstances |
| H2a: 'Doubled femininity' hypothesis | Predicts that Chinese American women will suffer more in heavily male fields because women of Chinese descent suffer from the burden of racialized feminization due to the stereotyping of Asianness as feminine |
| H3: Dishonesty stereotype and fraud guarding hypothesis | Predicts that scientists in fields with higher base rates of fraud will unconsciously guard against scientific dishonesty by discriminating against members of minority groups that are stereotyped as dishonest, including scholars of Chinese descent |
| H4: Creativity stereotype hypothesis | Predicts that because scientists claim to value creativity and because Chinese Americans are stereotyped as not creative, fields with lower degress of internal consensus (i.e. 'softer' sciences) will discriminate against Chinese scholars |
| H5: Attractiveness hypothesis | Predicts that Chinese women, who are seen as doubly feminine and also as generally attractive, may enjoy professional benefits due to the halo effect of attractiveness. Asian men, who are seen as feminine and also as less attractive, suffer instead from bias. |
| H6: Subordinate Male Target Hypothesis | Predicts that Chinese men suffer more than Chinese women in male-domianated fields because White men target non-White preferentially |

Table 1: A summary of hypotheses for bias against Chinese Americans in American science

|  | Non-Chinese | Chinese | % Chinese |
|---|---|---|---|
| Sociology | 1291 | 53 | 4% |
| Economics | 1895 | 192 | 9% |
| Physics | 3158 | 323 | 9% |
| Chemistry | 2953 | 359 | 11% |

Table 2: Chineseness by field

|  | Female | Male | % Male |
|---|---|---|---|
| Sociology | 680 | 664 | 49% |
| Economics | 257 | 1017 | 79% |
| Chemistry | 540 | 1674 | 75% |
| Physics | 336 | 2004 | 85% |

Table 3: Maleness by field

# FIGURES

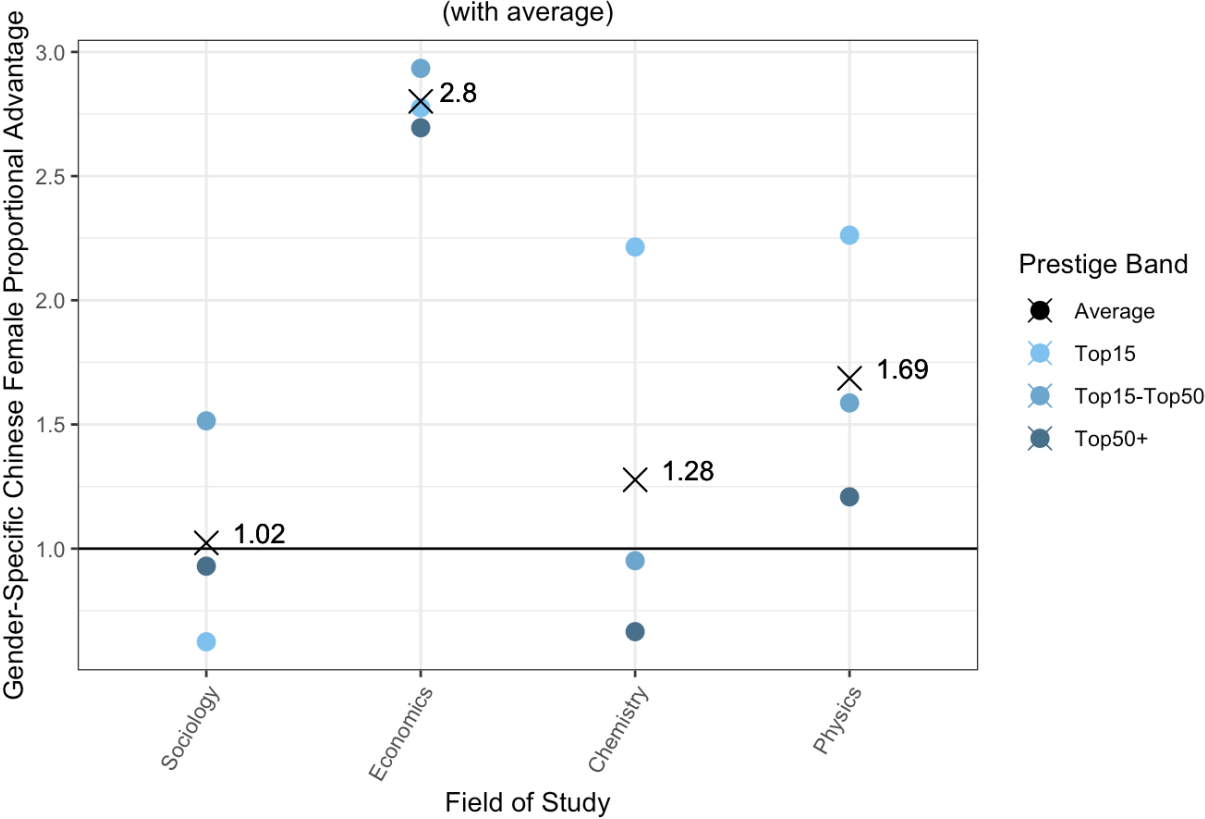

Figure 1: Gender-specific Chinese female proportional advantage by field

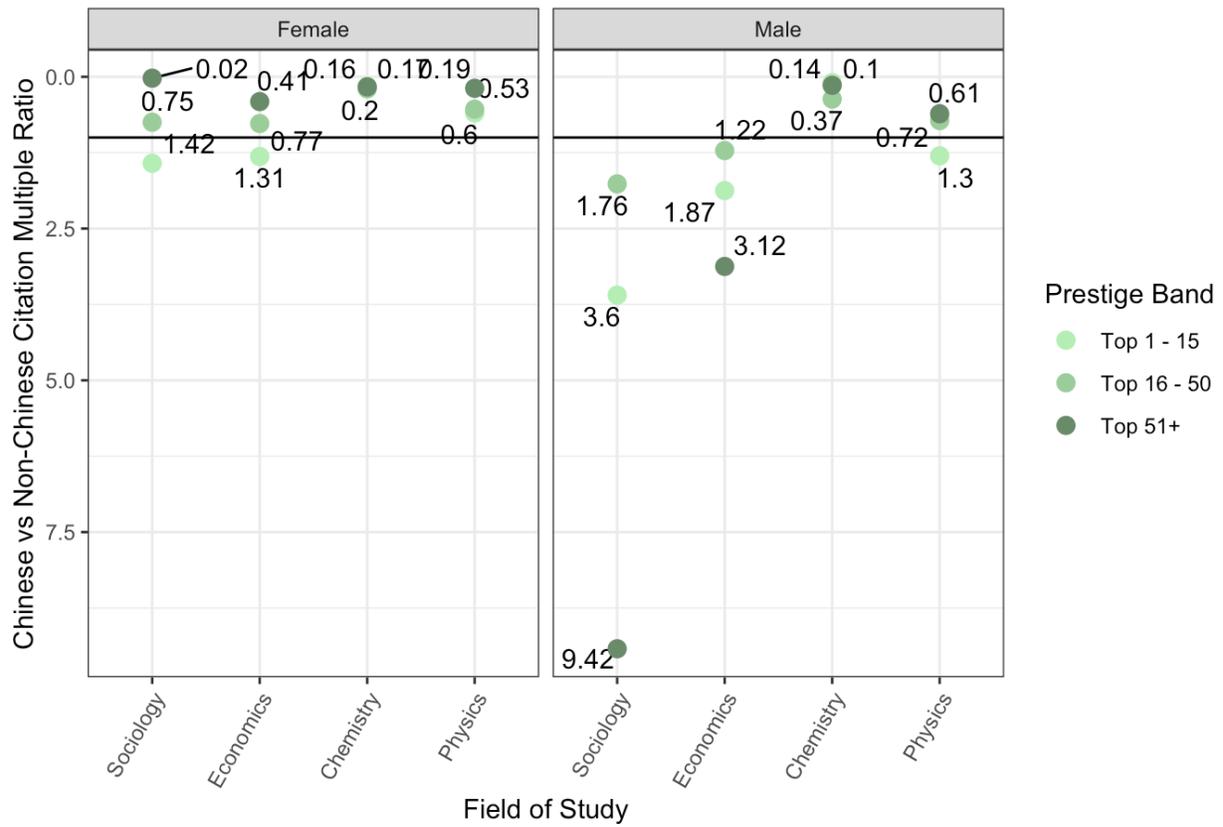

Figure 2: Chinese bias across 4 fields via ratio

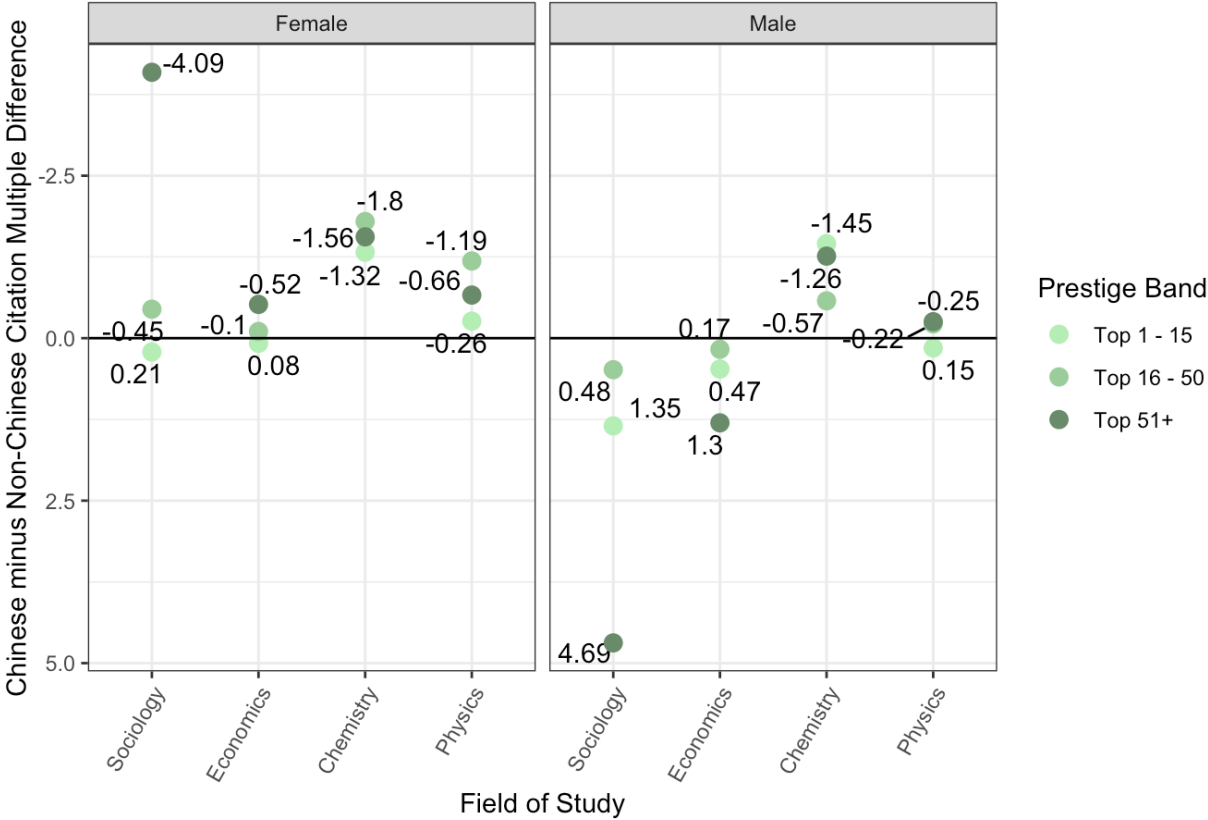

Figure 3: Chinese bias across 4 fields via difference

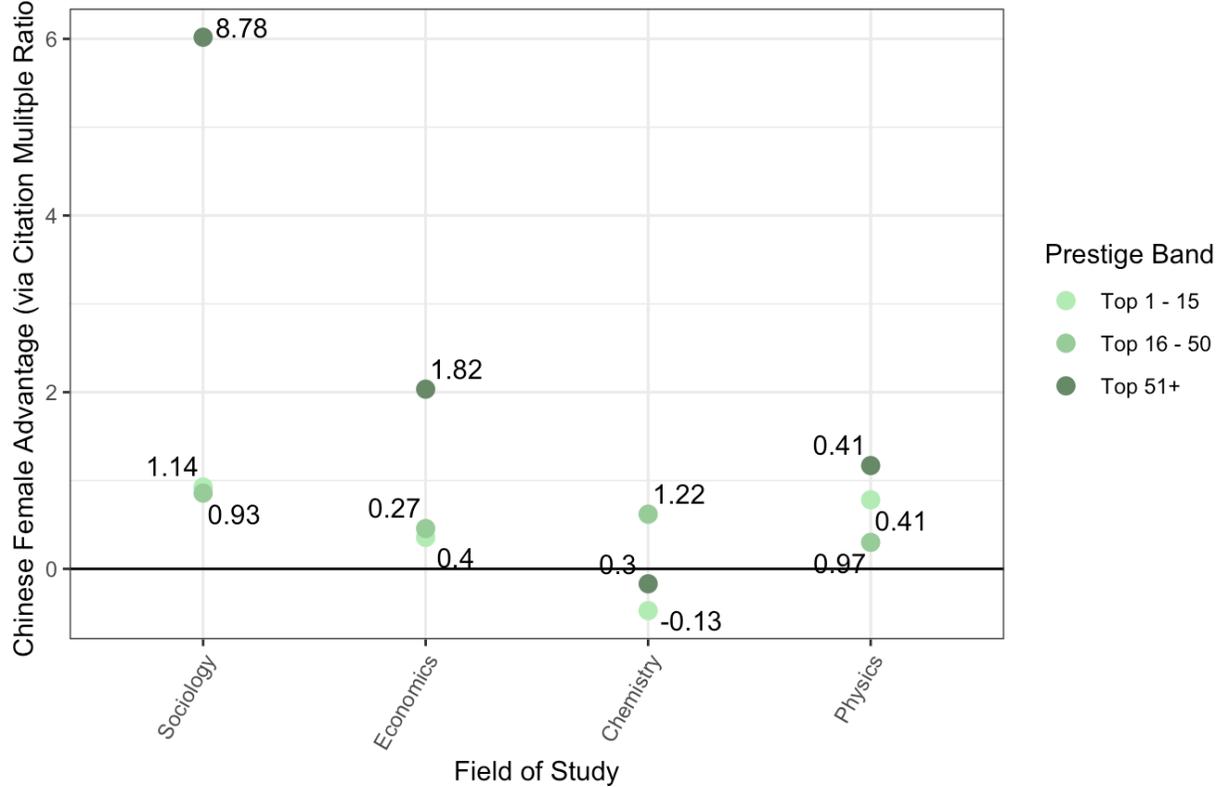

Figure 4: Chinese Female Advantage via ratio

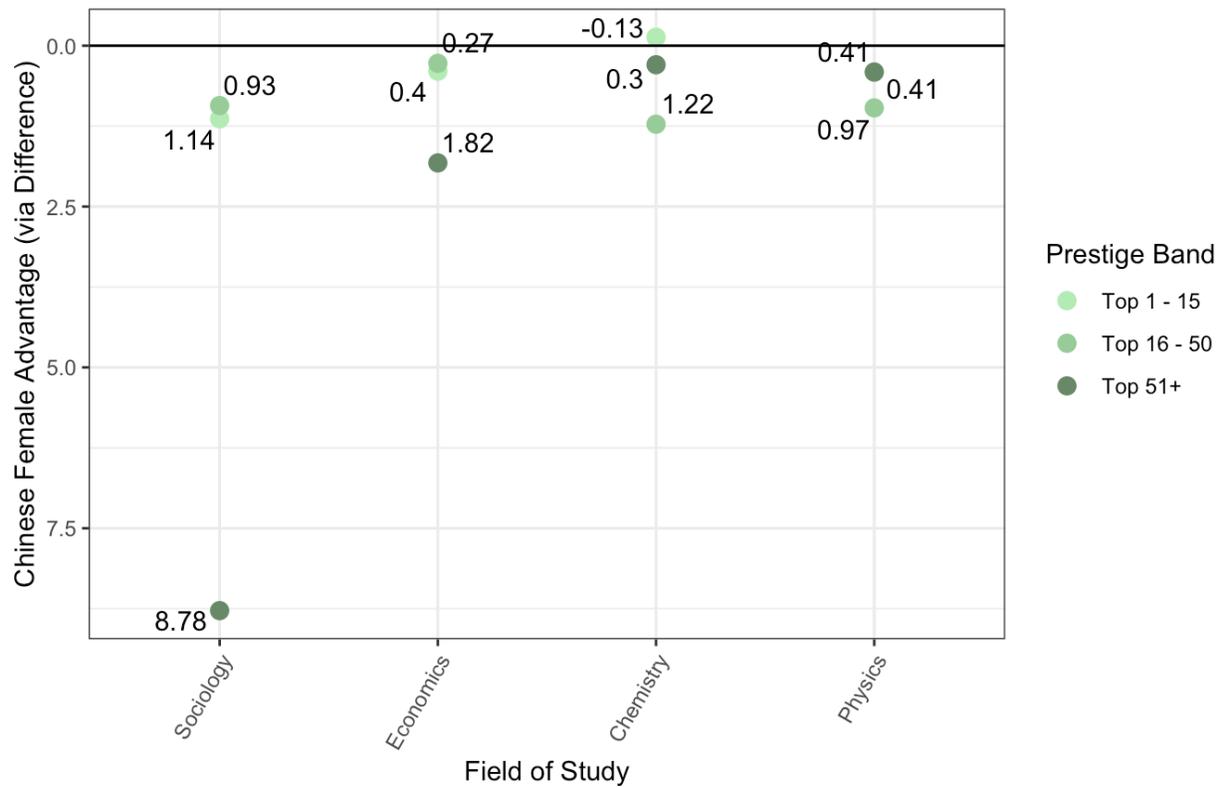

Figure 5: Chinese female advantage via difference

# SUPPLEMENTARY TABLES

| Prestige Band | Sociology | Economics | Chemistry | Physics |
|---|---|---|---|---|
| **Top 15** | 279 (24.9%) | 222 (23.9%) | 356 (21.7%) | 388 (27.3%) |
| **Top 15-50** | 513 (45.8%) | 374 (40.3%) | 573 (34.9%) | 551 (33.8%) |
| **Top 50+** | 328 (29.3%) | 332 (35.8%) | 715 (43.5%) | 481 (33.9%) |
| **Total** | 1120 | 928 | 1644 | 1420 |

Table S1: Professorial headcount by field and prestige band

**SUPPLEMENTARY FIGURES**

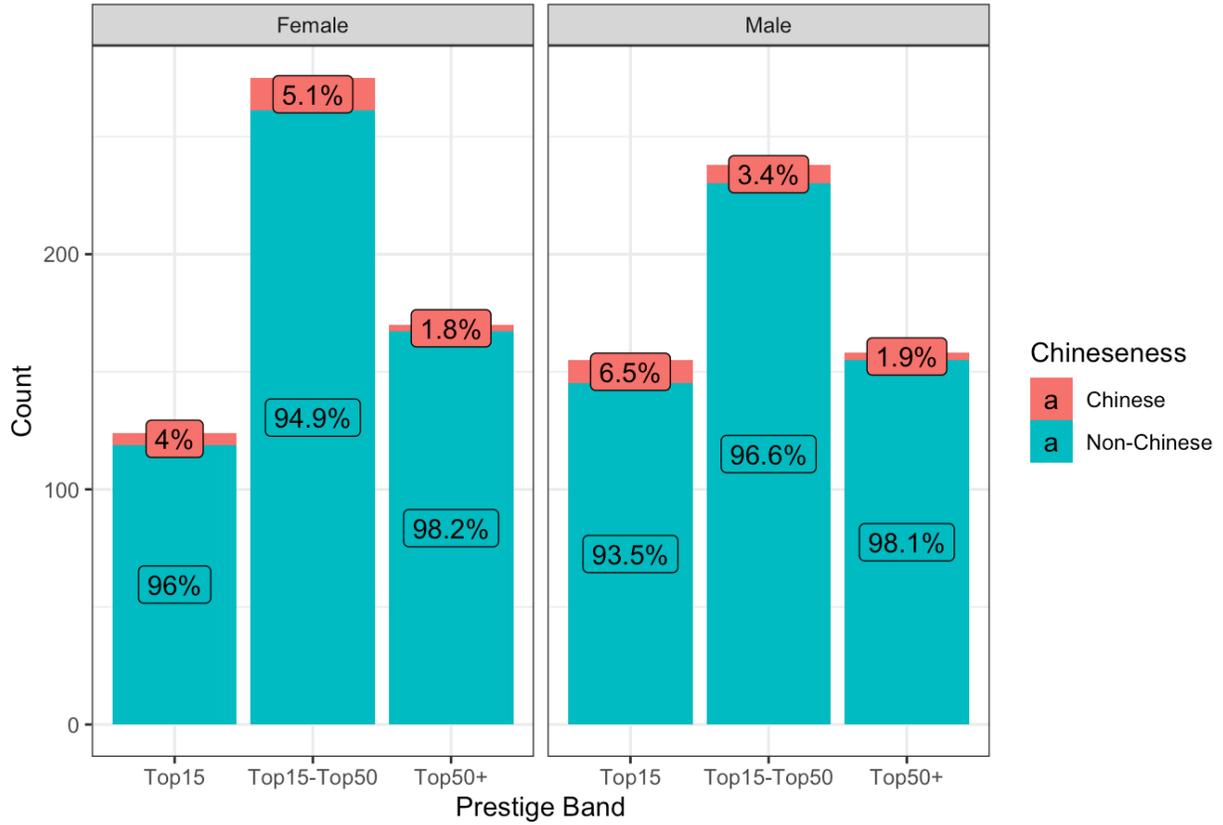

Figure S1: Proportion of Chinese and non-Chinese professors by gender in sociology

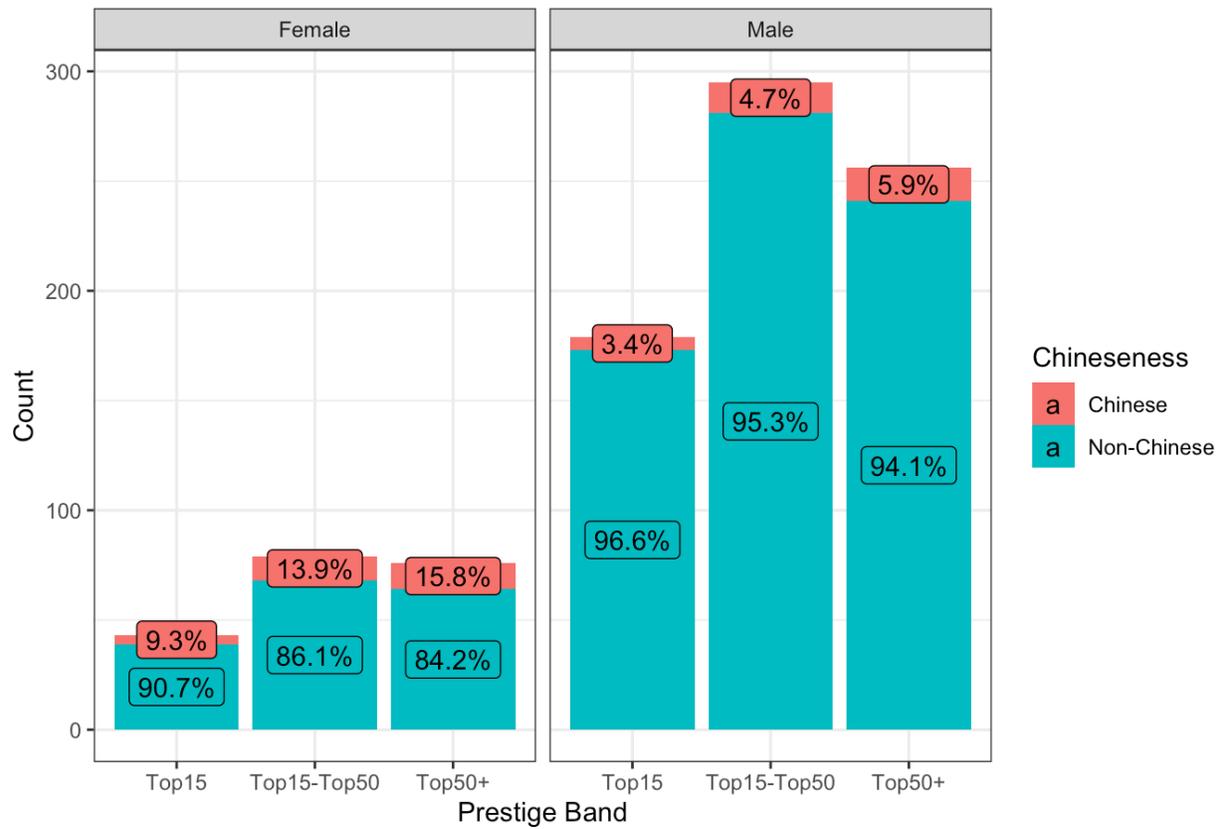

Figure S2: Proportion of Chinese and non-Chinese professors by gender in economics

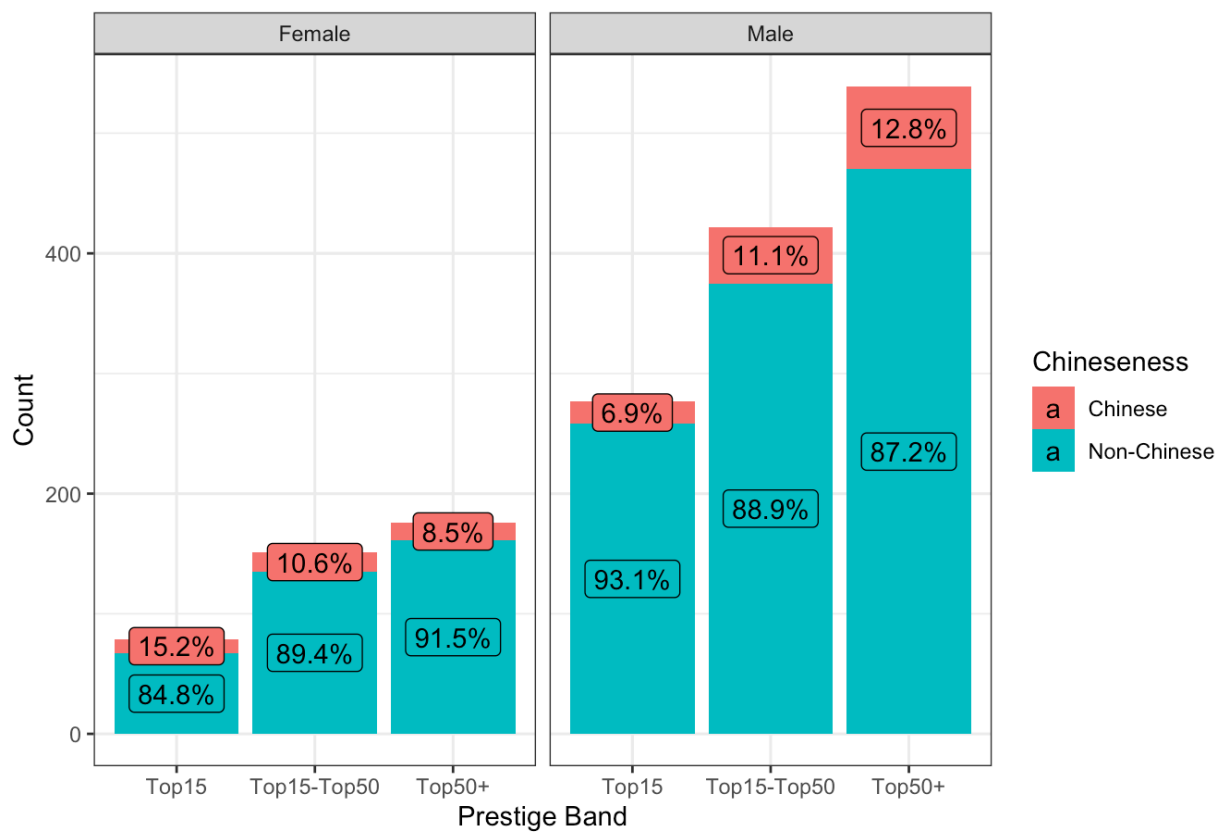

Figure S3: Proportion of Chinese and non-Chinese professors by gender in chemistry

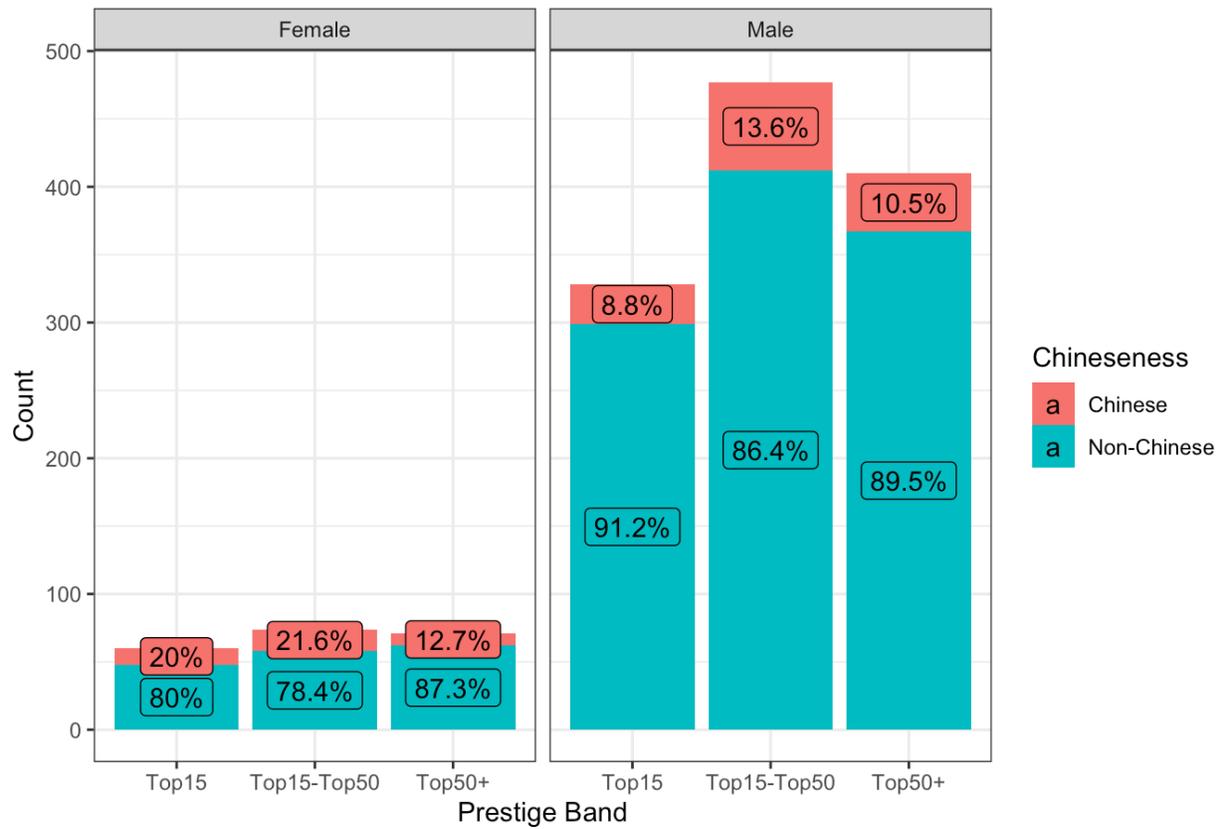

Figure S4: Proportion of Chinese and non-Chinese professors by gender in physics

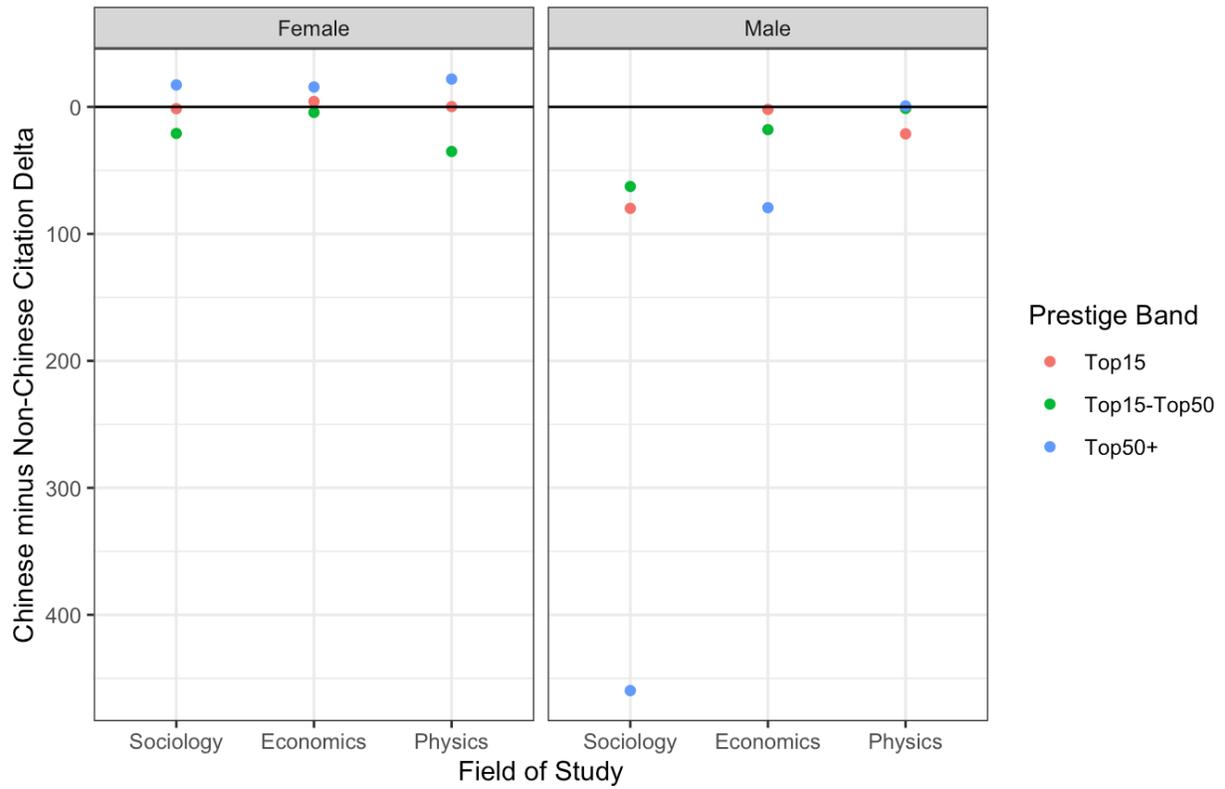

Figure S5: Robustness analyses – optimal matching